\documentclass[prc,showpacs,preprintnumbers,superscriptaddress,floatfix]{revtex4}
\usepackage{amsfonts}
\usepackage{graphicx}

\begin{document}

\title{Scattering of a Klein-Gordon particle by a Hulth\'en potential}
\author{Jian You Guo}
\email{jianyou@ahu.edu.cn}
\affiliation{School of physics and material science, Anhui university, Hefei 230039, P.R.
China}
\author{Xiang Zheng Fang}
\affiliation{School of physics and material science, Anhui university, Hefei 230039, P.R.
China}
\author{Chuan Mei Xie}
\affiliation{School of physics and material science, Anhui university, Hefei 230039, P.R.
China}

\begin{abstract}
The Klein-Gordon equation in the presence of a spatially one-dimensional
Hulth\'en potential is solved exactly and the scattering solutions are
obtained in terms of hypergeometric functions. The transmission coefficient
is derived by the matching conditions on the wavefunctions and the condition
for the existence of transmission resonances are investigated. It is shown
how the zero-reflection condition depends on the shape of the potential.
\end{abstract}

\maketitle

The study of low-momentum scattering in the Schr\"oinger equation in
one-dimensional even potentials shows that, as momentum goes to zero, the
reflection coefficient goes to unity unless the potential $V(x)$ supports a
zero-energy resonance\cite{Newton82}. In this case the transmission
coefficient goes to unity, becoming a transmission resonance\cite{Bohm51}.
Recently, this result has been generalized to the Dirac equation\cite%
{Dombey00}, showing that transmission resonances at $k=0$ in the Dirac
equation take place for a potential barrier $V=V(x)$ when the corresponding
potential well $V=-V(x)$ supports a supercritical state. This conclusion is
demonstrated in both special examples as square potential and Gaussian
potential, where the phenomenon of transmission resonance is exhibited
clearly in Dirac spinors in the appropriate shapes and strengths of the
potentials. Except for the both special examples, the transmission resonance
is also investigated in the realistic physical system. In Ref.\cite%
{Kennedy02}, a key potential in nuclear physics is introduced, and the
scattering and bound states are obtained by solving the Dirac equation in
the presence of Woods-Saxon potential, which has been extensively discussed
in the literature\cite{Guo02,Petri03,Chen04,Guo03,Alhai01}. The transmission
resonance is shown appearing at the spinor wave solutions with a functional
dependence on the shape and strength of the potential. The presence of
transmission resonance in relativistic scalar wave equations in the
potential is also investigated by solving the one-dimensional Klein-Gordon
equation. The phenomenon of resonance appearing in Dirac equation is
reproduced at the one-dimensional scalar wave solutions with a functional
dependence on the shape and strength of the potential similar to those
obtained for the Dirac equation\cite{Rojas05}.

Due to the transmission resonance appearing in the realistic physical system
for not only Dirac particle but also Klein Gordon particle as illustrated in
the Woods-Saxon potential, it is indispensable to check the existence of the
phenomenon in some other fields. Considering that the Hulth\'{e}n potential%
\cite{Hulth42} is an important realistic model, it has been widely used in a
number of areas such as nuclear and particle physics, atomic physics,
condensed matter and chemical physics\cite{Varsh90,Jamee86,Barna87,Richa92}.
Hence, to discuss the scattering problem for a relativistic particle moving
in the potential is significant, which may provide more knowledge on the
transmission resonance. Recently, there have been a great deal of works to
be put to the Hulth\'{e}n potential in order to obtain the bound and
scattering solutions in the case of relativity and non relativity\cite%
{Chen07}. However, the transmission resonance is not still checked for
particle moving in the potential in the relativistic case. In this paper, we
will derive the scattering solution of the Klein-Gordon equation in the
presence of the general Hulth\'{e}n potential, and show the phenomenon of
transmission resonance as well as its relation to the parameters of the
potential.

Following Ref.\cite{Rojas05}, one-dimensional Klein-Gordon equation,
minimally coupled to a vector potential $A^{\mu }$, is written as
\begin{equation}
\eta ^{\alpha \beta }\left( \partial _{\alpha }+ieA_{\alpha }\right) \left(
\partial _{\beta }+ieA_{\beta }\right) \phi +\phi =0,
\end{equation}%
where the metric $\eta ^{\alpha \beta }=$diag(1,-1). For simplicity, the
natural units $\hbar =c=m=1$ are adopted, and Eq.(1) is simplified into the
following form
\begin{equation}
\frac{d^{2}\phi (x)}{dx^{2}}+\left\{ \left[ E-V(x)\right] ^{2}-1\right\}
\phi (x)=0.
\end{equation}%
In Eq.(2), $V(x)$ is chosen as the general Hulth\'{e}n potential with the
definition\cite{Hulth42,Qiang07} as%
\begin{equation}
V(x)=\Theta (-x)\frac{V_{0}}{e^{-ax}-q}+\Theta (x)\frac{V_{0}}{e^{ax}-q},
\end{equation}%
where all the parameters $V_{0}$, $a$, and $q$ are real and positive. To
remove off the divergence of Hulth\'{e}n potential, $q<1$ is required. If $%
q=-1$ is taken, the Hulth\'{e}n potential turns into a Woods-Saxon
potential. $\Theta (x)$ is the Heaviside step function. The form of the Hulth%
\'{e}n potential is shown in Fig.1 and 2 at different values of parameters.

From Fig.1 and 2 one readily notices that for a given value of the potential
strength parameter $V_{0}$, as $q$ increases, the height of potential
barrier increases. When $q\longrightarrow 1$, the height of potential
barrier goes to infinity. Similarly, the potential becomes more diffusible
with the decreasing of the diffuseness parameter $a$.

In order to obtain the scattering solutions for $x<0$ with $E^{2}>1$, we
solve the differential equation%
\begin{equation}
\frac{d^{2}\phi (x)}{dx^{2}}+\left\{ \left[ E-\frac{V_{0}}{e^{-ax}-q}\right]
^{2}-1\right\} \phi (x)=0.
\end{equation}%
On making the substitution $y=qe^{ax}$, Eq.(4) becomes%
\begin{equation}
a^{2}y^{2}\frac{d^{2}\phi }{dy^{2}}+a^{2}y\frac{d\phi }{dy}+\left[ \left( E-%
\frac{V_{0}}{q}\frac{y}{1-y}\right) ^{2}-1\right] \phi (x)=0.
\end{equation}%
In order to derive the solution of Eq.(5), we put $\phi =y^{\mu }\left(
1-y\right) ^{\lambda }f$, then Eq.(5) reduces to the hypergeometric equation

\begin{eqnarray}
&&y\left( 1-y\right) f^{\prime \prime }+\left[ 1+2\mu -\left( 2\mu +2\lambda
+1\right) y\right] f^{\prime }  \nonumber \\
&&-\left( \lambda (1+2\mu )+\frac{2EV_{0}}{a^{2}q}\right) f=0,
\end{eqnarray}%
where the primes denote derivatives with respect to $y$ and the parameters $%
\mu $, $k$, $\lambda _{\pm }$, $\nu $ are
\begin{eqnarray*}
\mu  &=&ik/a,\text{ \ }k=\sqrt{E^{2}-1}, \\
\lambda _{\pm } &=&\frac{1}{2}\pm \frac{1}{2}\sqrt{1-\left( 2V_{0}/aq\right)
^{2}}, \\
\nu  &=&\sqrt{\mu ^{2}+\lambda ^{2}-\lambda -\frac{2EV_{0}}{a^{2}q}}.
\end{eqnarray*}%
The general solution of Eq.(6) can be expressed in terms of hypergeometric
function as
\begin{eqnarray}
f(y) &=&A\text{ }F\left( \mu -\nu +\lambda ,\mu +\nu +\lambda ,1+2\mu
;y\right)   \nonumber \\
&&+B\text{ }y^{-2\mu }F\left( -\mu -\nu +\lambda ,-\mu +\nu +\lambda ,1-2\mu
;y\right) .
\end{eqnarray}%
So
\begin{eqnarray}
\phi _{L}(y) &=&A\text{ }y^{\mu }\left( 1-y\right) ^{\lambda }F\left( \mu
-\nu +\lambda ,\mu +\nu +\lambda ,1+2\mu ;y\right)   \nonumber \\
&&+B\text{ }y^{-\mu }\left( 1-y\right) ^{\lambda }F\left( -\mu -\nu +\lambda
,-\mu +\nu +\lambda ,1-2\mu ;y\right) .
\end{eqnarray}%
As $x\longrightarrow -\infty $, there is $y\longrightarrow 0$. So, the
asymptotic behavior of $\phi _{L}(y)$ can be written as
\begin{equation}
\lim_{x\longrightarrow -\infty }\phi
_{L}(y)=Aq^{ik/a}e^{ikx}+Bq^{-ik/a}e^{-ikx}.
\end{equation}

Next, we consider the solution of Eq.(2) for $x>0$. With the potential
represented in Eq.(3), the differential equation to solve becomes
\begin{equation}
\frac{d^{2}\phi (x)}{dx^{2}}+\left\{ \left[ E-\frac{V_{0}}{e^{ax}-q}\right]
^{2}-1\right\} \phi (x)=0.
\end{equation}

The analysis of the solution can be simplified making the substitution $%
z=qe^{-ax}$. Eq.(10) can be written as%
\begin{equation}
a^{2}z^{2}\frac{d^{2}\phi }{dz^{2}}+a^{2}z\frac{d\phi }{dz}+\left\{ \left[ E-%
\frac{V_{0}z}{q\left( 1-z\right) }\right] ^{2}-1\right\} \phi (x)=0.
\end{equation}%
Put $\phi =z^{\mu }\left( 1-z\right) ^{\lambda }g$, Eq.(11) reduces to the
hypergeometric equation

\begin{eqnarray}
&&z\left( 1-z\right) g^{\prime \prime }+\left[ 1+2\mu -\left( 2\mu +2\lambda
+1\right) z\right] g^{\prime }  \nonumber \\
&&-\left[ \lambda (1+2\mu )+\frac{2EV_{0}}{a^{2}q}\right] g=0,
\end{eqnarray}%
where the primes denote derivatives with respect to $z$. The general
solution of Eq.(12) is
\begin{eqnarray}
g(z) &=&C\left( \mu -\nu +\lambda ,\mu +\nu +\lambda ,1+2\mu ;z\right)
\nonumber \\
&&+Dz^{-2\mu }F\left( -\mu -\nu +\lambda ,-\mu +\nu +\lambda ,1-2\mu
;z\right) .
\end{eqnarray}%
So,
\begin{eqnarray}
\phi _{R}(z) &=&Cz^{\mu }\left( 1-z\right) ^{\lambda }F\left( \mu -\nu
+\lambda ,\mu +\nu +\lambda ,1+2\mu ;z\right)  \nonumber \\
&&+Dz^{-\mu }\left( 1-z\right) ^{\lambda }F\left( -\mu -\nu +\lambda ,-\mu
+\nu +\lambda ,1-2\mu ;z\right) .
\end{eqnarray}%
Keeping only the solution for the transmitted wave, $C=0$ in Eq.(14). As $%
x\longrightarrow +\infty (z\longrightarrow 0)$, there is
\begin{equation}
\phi _{R}(x)\longrightarrow Dq^{-ik/a}e^{ikx}.
\end{equation}%
The electrical current density for the one-dimensional Klein-Gordon equation
is given by the expression
\begin{equation}
J=\frac{i\hbar }{2m}\left( \phi \frac{d\phi ^{\ast }}{dx}-\phi ^{\ast }\frac{%
d\phi }{dx}\right) .
\end{equation}%
The current as $x\longrightarrow -\infty $ can be decomposed as $j_{L}=j_{%
\text{in}}-j_{\text{refl}}$ where $j_{\text{in}}$ is the incident current
and $j_{\text{refl}}$ is the reflected one. Analogously we have that, on the
right side, as $x\longrightarrow \infty $ the current is $j_{R}=j_{\text{%
trans}}$, where $j_{\text{trans}}$ is the transmitted current. Using the
reflected $j_{\text{refl}}$ and transmitted $j_{\text{trans}}$ currents, we
have that the reflection coefficient $R$, and the transmission coefficient $%
T $ can be expressed in terms of the coefficients $A$, $B$, and $D$ as
\begin{equation}
R=\frac{j_{\text{refl}}}{j_{\text{in}}}=\frac{\left\vert B\right\vert ^{2}}{%
\left\vert A\right\vert ^{2}},
\end{equation}%
\begin{equation}
T=\frac{j_{\text{trans}}}{j_{\text{in}}}=\frac{\left\vert D\right\vert ^{2}}{%
\left\vert A\right\vert ^{2}}.
\end{equation}

Obviously, $R$ and $T$ are not independent; they are related via the
unitarity condition
\begin{equation}
R+T=1.
\end{equation}
In order to obtain $R$ and $T$ we proceed to equate at $x=0$ the right $\phi
_{_{R}}$ and left $\phi _{_{L}}$ wave functions and their first derivatives.
From the matching condition we derive a system of equations governing the
dependence of coefficients $A$ and $B$ on $D$ that can be solved numerically.

The calculated transmission coefficient $T$ varying with the energy $E$ is
displayed in Figs.3-6 at the different values of the parameters in the
Hulth\'en potential. From Figs.3-6, one can see that the transmission
resonance appears in all the Hulth\'en potential considered here. But the
intensity and width of resonance as well as the condition for the existence
of resonance are different, and they depend on the shape of the potential.
Compared Fig.3 with Fig.4, it can be seen that the width of resonance
decreases as the decreasing of diffuseness $a$, which is similar to that of
Woods-Saxon potential as shown in Figs.3 and 5 in Ref.\cite{Rojas05}. The
same dependence can also be observed from Figs.5 and 6. Compared Fig.3 with
Fig.5, one can find that the condition for the existence of transmission
resonance does also relate to the parameter $q$. As $q$ decreases, the
height of potential barrier increases, the widths of the transmission
resonance increases. The conclusion can also be obtained by comparing Fig.4
with Fig.6. In order to obtain more knowledge on the dependence of
transmission resonance on the shapes of the potential, the transmission
coefficient $T$ varying with the strength of potential $V_0$ is plotted in
Figs.7 and 8. Beside of the phenomenon of transmission resonance, similar to
the Fig.3 and 4, the width of resonance decreasing as the decreasing of
diffuseness $a$ is disclosed. All these show the transmission resonances in
Hulth\'en potential for Klein-Gordon particle possess the same rich
structure with that we observe in Woods-Saxon potential.

\begin{acknowledgments}
This work was partly supported by the National Natural Science Foundation of
China under Grant No. 10475001 and 10675001, the Program for New Century
Excellent Talents in University of China under Grant No. NCET-05-0558, the
Program for Excellent Talents in Anhui Province University, and the
Education Committee Foundation of Anhui Province under Grant No. 2006KJ259B
\end{acknowledgments}

\begin{figure}[!h]
\centering \includegraphics[width=8.cm]{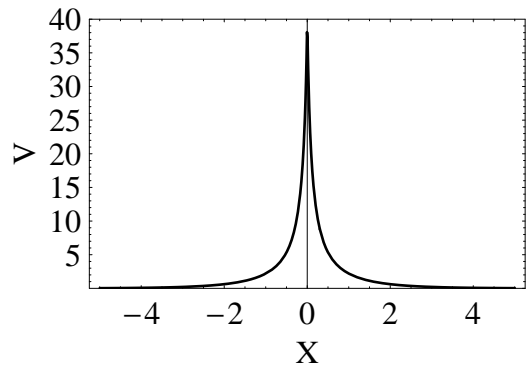} \vspace{-10pt}
\caption{Hulth\'en potential for a=1.0 and q=0.9 with V$_0$=4, of which the
peak of barrier reaches 40.0.}
\end{figure}

\begin{figure}[!h]
\centering \includegraphics[width=8.0cm]{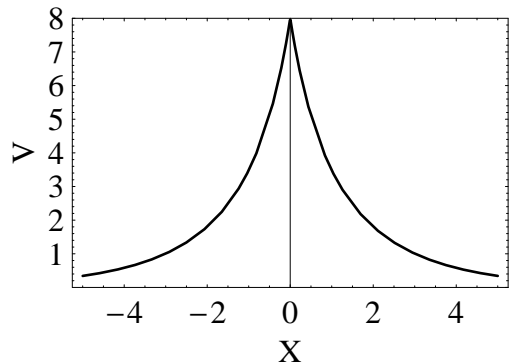} \vspace{-10pt}
\caption{Hulth\'en potential for a=0.5 and q=0.5 with V$_0$=4, of which the
peak of barrier reaches 8.0.}
\end{figure}

\begin{figure}[!h]
\centering \includegraphics[width=8.cm]{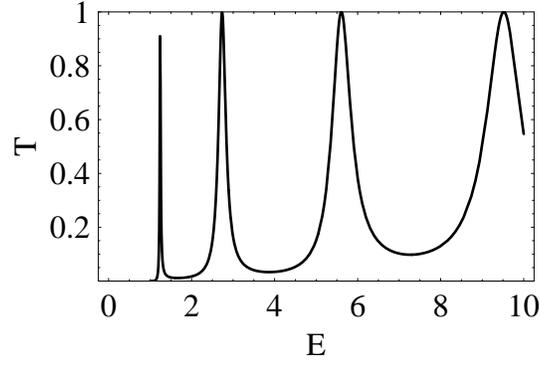} \vspace{-10pt}
\caption{The transmission coefficient for the relativistic Hulth\'en
potential barrier. The plot illustrate $T$ for varying energy $E$ with $%
V_0=4,a=1$, and $q=0.9$.}
\end{figure}
\begin{figure}[!h]
\centering \includegraphics[width=8.0cm]{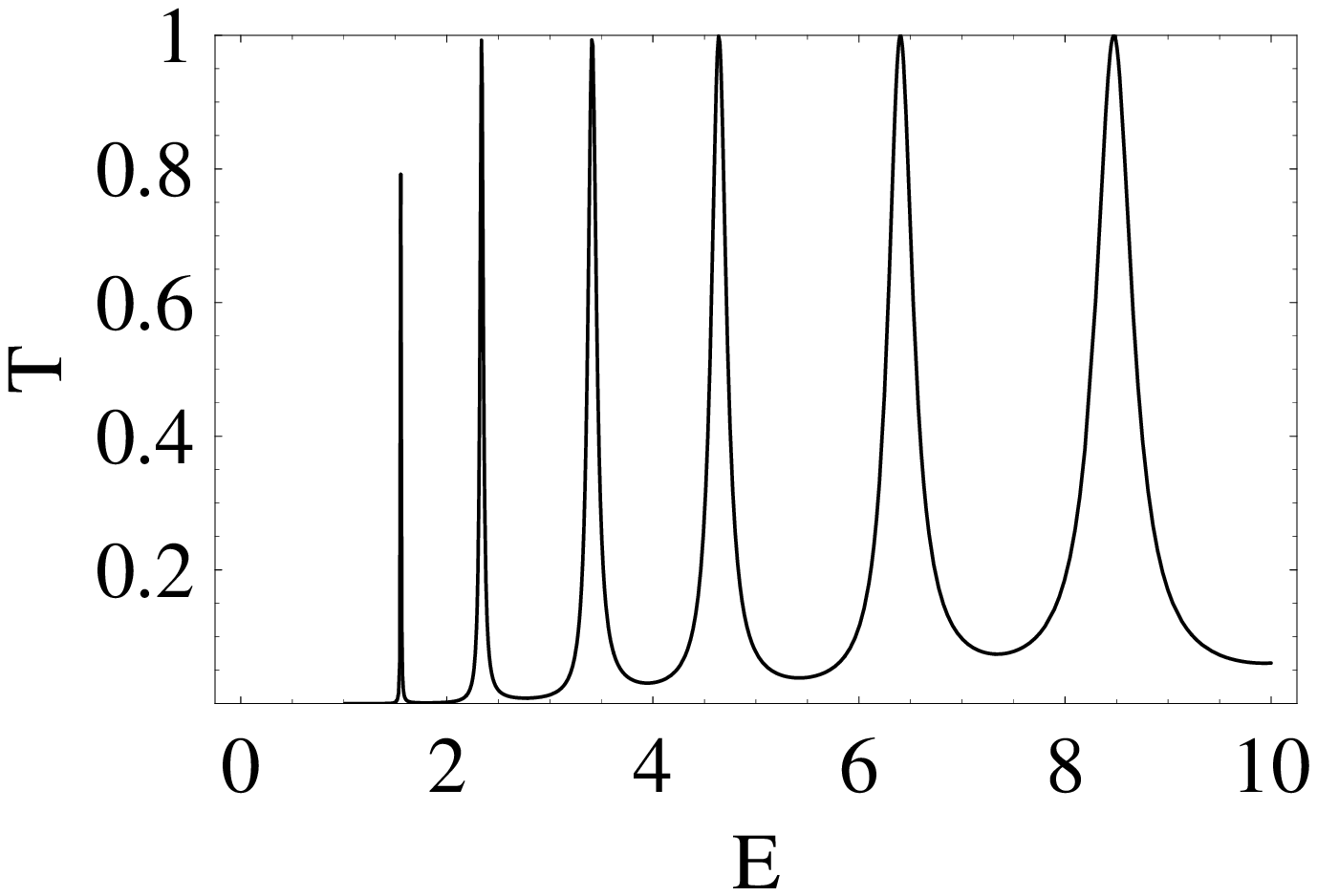} \vspace{-10pt}
\caption{Similar to Fig.3, but with $V_0=4,a=0.5$, and $q=0.9$.}
\end{figure}

\begin{figure}[!h]
\centering \includegraphics[width=8.cm]{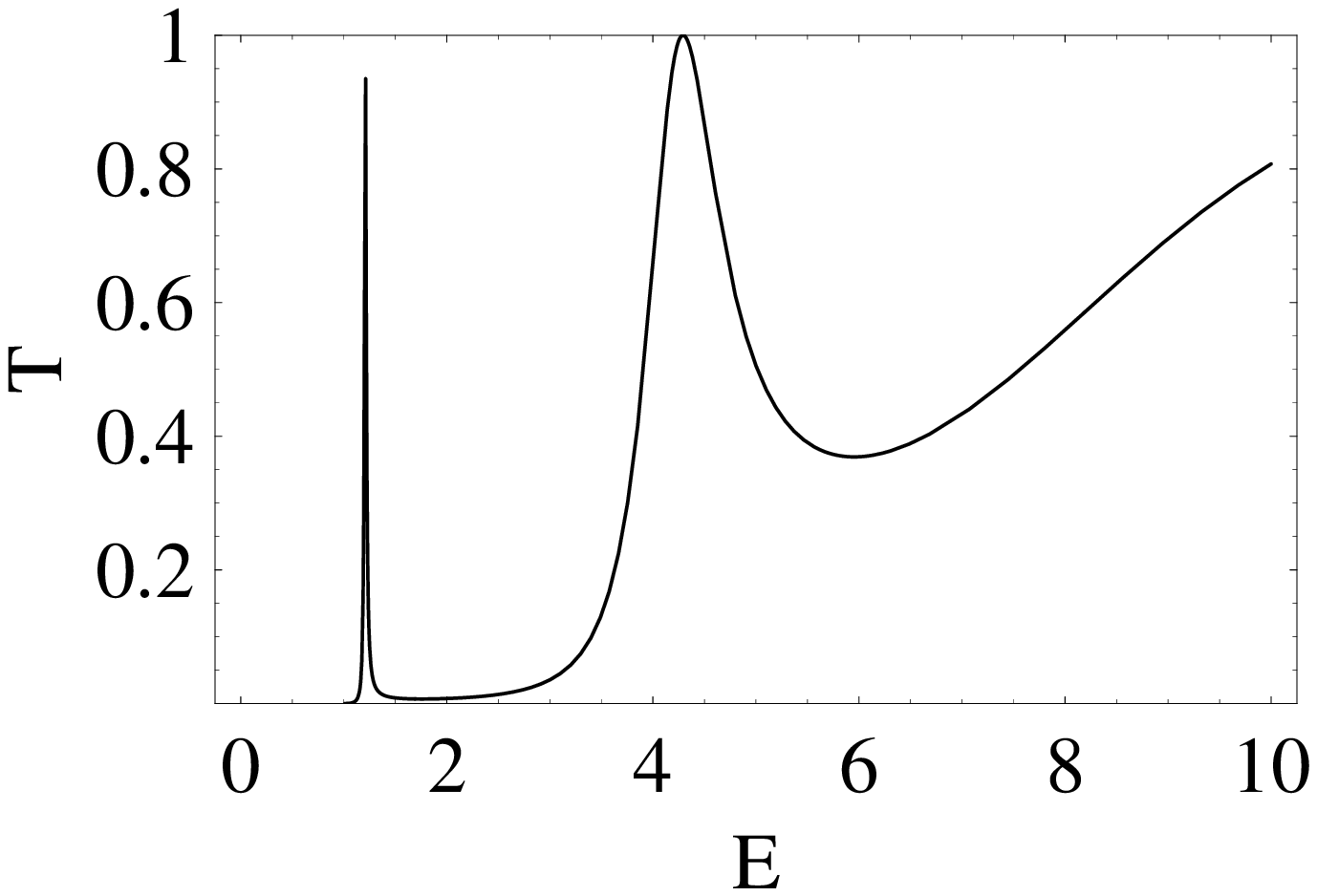} \vspace{-10pt}
\caption{Similar to Fig.3, but with $V_0=4,a=1$, and $q=0.5$.}
\end{figure}
\begin{figure}[!h]
\centering \includegraphics[width=8.0cm]{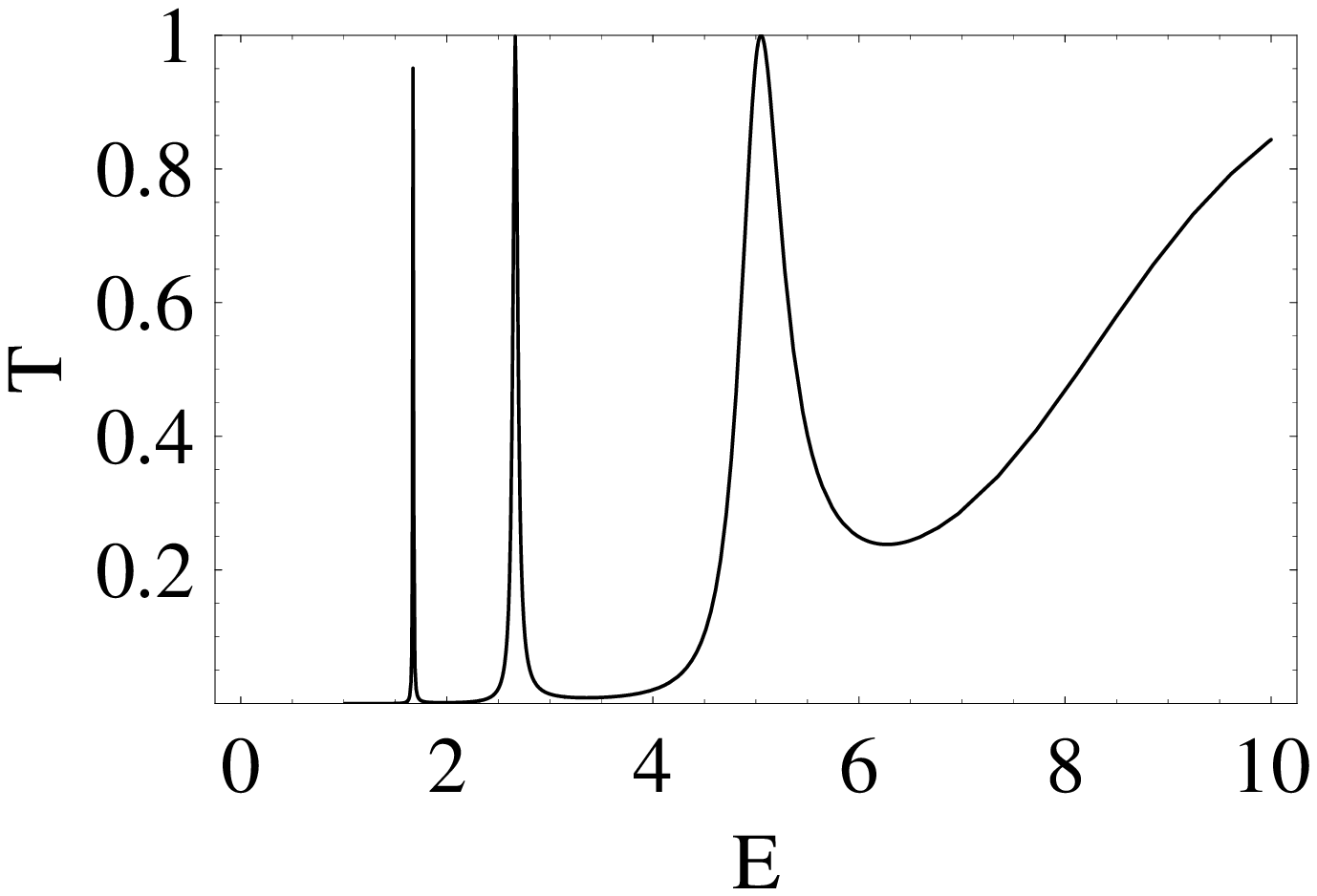} \vspace{-10pt}
\caption{Similar to Fig.3, but with $V_0=4,a=0.5$, and $q=0.5$.}
\end{figure}

\begin{figure}[!h]
\centering \includegraphics[width=8.cm]{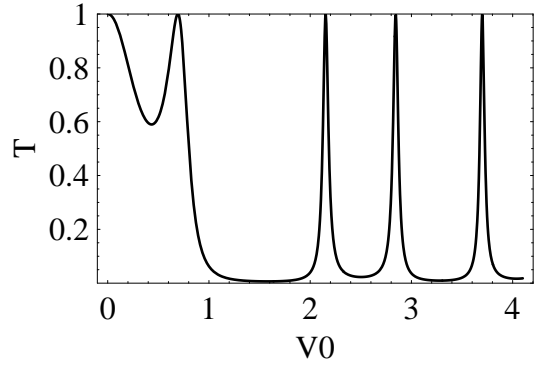} \vspace{-10pt}
\caption{The transmission coefficient for the relativistic Hulth\'en
potential barrier. The plot illustrate $T$ for varying barrier height $V_0$
with $E=2,a=1$, and $q=0.9$.}
\end{figure}
\begin{figure}[!h]
\centering \includegraphics[width=8.0cm]{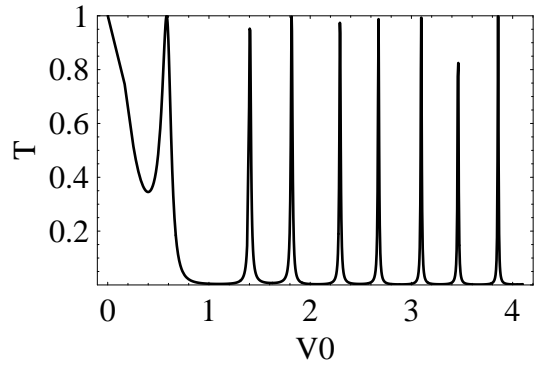} \vspace{-10pt}
\caption{Similar to Fig.7, but with $E=2,a=0.5$, and $q=0.9$.}
\end{figure}

\end{document}